\documentclass[twocolumn,twoside,slac_two]{revtex4}
\usepackage{graphicx}
\usepackage{fancyhdr}
\usepackage{hyperref}
\hypersetup{
    colorlinks=true,
    urlcolor=magenta,
}
\begin{document}

\title{Spatial and temporal variations of high-energy electron flux in the outer radiation belt}


\author{S.V, Koldashov. S.Yu. Aleksandrin, N.D. Eremina}
\affiliation{National Research Nuclear University MEPhI (Moscow Engineering Physics Institute)}

\begin{abstract}
The results of observation of short-term variations of high-energy electron flux in the outer radiation belt, obtained in ARINA satellite experiment  (2006 - 2016), are presented. Scintillation spectrometer ARINA on board the Resurs-DK1 Russian satellite has been developed in MEPhI. The instrument carried out continuous measurements of high-energy electron flux and its energy spectrum in low-Earth orbits in the range 3-30 MeV with 10 – 15\% energy resolution. A time profile of electron flux at different L - shells has been studied in detail on the example of March 2012, and analysis of experimental data on high-energy (4-6 MeV) electrons in the outer radiation belt zone (L$\sim$ 3–-7) was fulfilled. It was shown a large variability of flux of such electrons there. The sharp effects in electron flux (as rise and as fall) in magnetosphere interrelated with geomagnetic storms caused by solar flares and coronal mass ejections have been observed.   
\end{abstract}

\maketitle

\thispagestyle{fancy}


\section{Introduction}
As it’s well known from earlier satellite experiments, the outer radiation belt is very high dynamic formation \cite{bib:bibl1}. Some mechanisms of relativistic electron belt generation and loss are considered in \cite{bib:bibl2, bib:bibl3}.

Latest measurements of relativistic electron flux in the outer radiation belt, fulfilled by two Van-Allen probes in high apogee orbit, gave new very important information about the behavior of high-energy electron ensemble in the outer belt \cite{bib:bibl4}. On the basis of this observation it is asserted that there is the local structure of outer radiation belt with a large variability. In particular it was noted that the third radiation belt was formed in the outer magnetosphere at the beginning of September 2012.

It is also necessary to mention the latest measurements of outer belt electron flux by PROBA-V satellite in the low Earth polar orbit since May 2013. Deep dropout of electron flux at L$>$4 was observed by EPT instrument onboard this satellite during the main phase of big geomagnetic storm event on 17 March 2015 \cite{bib:bibl5}.

Russian satellite ARINA experiment, aimed to study fast variations of high-energy electron and proton fluxes in the magnetosphere, was carried out in low Earth orbit since 2006 till 2016 \cite{bib:bibl6}. Detail results concerning the observation of high-energy electron flux in zone of outer radiation belt at L shell range of 2.5 –- 7.0 in March 2012 are presented below.

\section{Instrument and experiment}
The ARINA scintillation spectrometer developed by the MEPhI detects and identifies electrons (3-30 MeV) and protons (30-100 MeV), measures particle energies, and allows to study the energy spectra and time profiles of particle fluxes.
The ARINA experiments was carried out on board the low-orbital Resurs-DK1 satellite \cite{bib:bibl6} with altitude of 350-600 km and an orbit inclination of 70$^{o}$. The experiment was executed since mid-June 2006 till January 2016.
Local pitch-angles of particles during the measurements were closed to 90$^{o}$.
The multilayer scintillation detector (ten scintillation layers) is main part of the instrument. Charged particles (electrons, protons) move in the instrument aperture, defining by three first segmented layers, sequentially pass through scintillation layers, lose energy, and are absorbed in the detector. Particles passed through the entire instrument are cut by last bottom layer of detector operating in the anticoincidence mode. Thus, particles stopped in the multilayer detector are electrons with energies of 3-30 MeV and protons with energies of 30-100 MeV. Particles are identified by the energy release in each scintillation layer when they passing through the instrument in combination of particle range in detector layers. The electron and proton energies are measured by their range in detector. The physics scheme and performances of the ARINA instrument are described in detail in \cite{bib:bibl7}. 
The ARINA instrument give the possibility to measure the energy spectra of particles with 10\%-–15\% energy resolution and to trace spectra evolution, to determine time profiles of particle flux variations with high time resolution and can operate in high-intensity particle fluxes in the radiation belts or during power solar proton events The instrument acceptance controlled by the configuration and arrangement of three top detector layers is about 10 cm$^{2}$sr. At that instrument field of view is about ±20$^{o}$ and angular resolution within 7$^{o}$ –- 10$^{o}$. 
The main objective of experiments is the study of high-energy charged particle bursts of solar and geophysical origin in the inner zone of magnetosphere \cite{bib:bibl8}.
In this work we used the ARINA experimental data on the high-energy electron flux in the energy range of 4-6 MeV in the outer radiation belt.

\section{High-energy electron flux in the outer radiation belt}

It is known that geomagnetic disturbances are one of the most important factors, defining the behavior of outer radiation belt. Below we consider high-energy electron flux in zones of outer radiation belt and slot between inner and outer belts on example of one month and note main features in behavior of electron flux under different magnetospheric conditions. 

Fig. 1 presents Dst variation during March 2012.
Most large geomagnetic storm on 9-10 March  led to formation long-lived high-energy electron belt in zone of outer radiation belt. The slot region at L=2.5-2.7 (Fig. 2) is usually insensitive to magnetospheric disturbances (even large one). But sometimes under specific conditions this region  can be also fill by high-energy electrons forming long-lived belts \cite{bib:bibl1}.
Fig. 3 demonstrates period of formation high-energy electron belt with flux 102 times higher than background albedo electron flux in inner region (L=2.9 - 3.1) of outer radiation belt during the recovery phase of geomagnetic storm on 9-10 March 2012.

\begin{figure} [b]
\includegraphics[width=85mm]{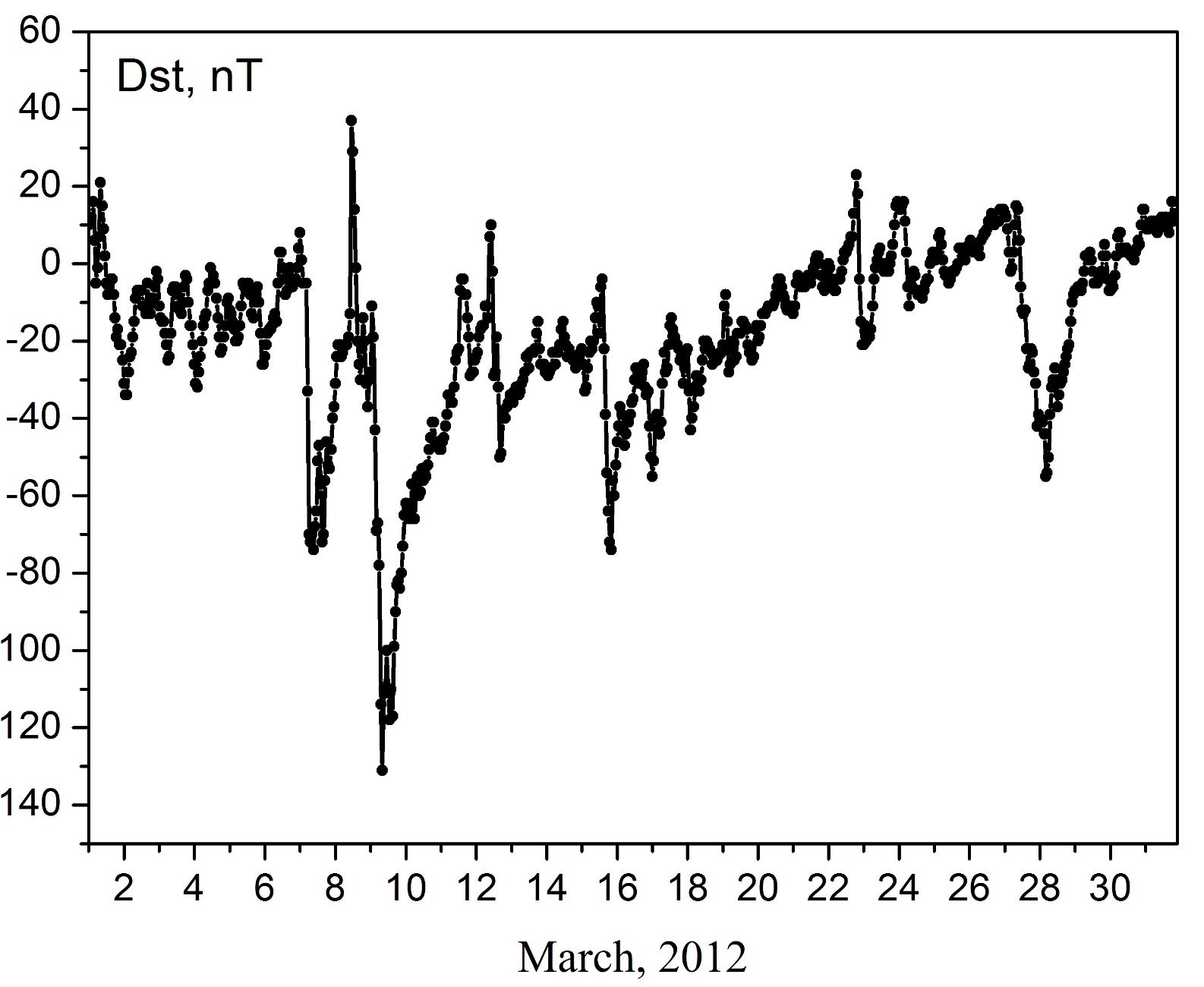}
\caption{Dst variation during March 2012.
http:$//$wdc.kugi.kyoto-u.ac.jp}
\label{fig1}
\end{figure}

\begin{figure}
\includegraphics[width=85mm]{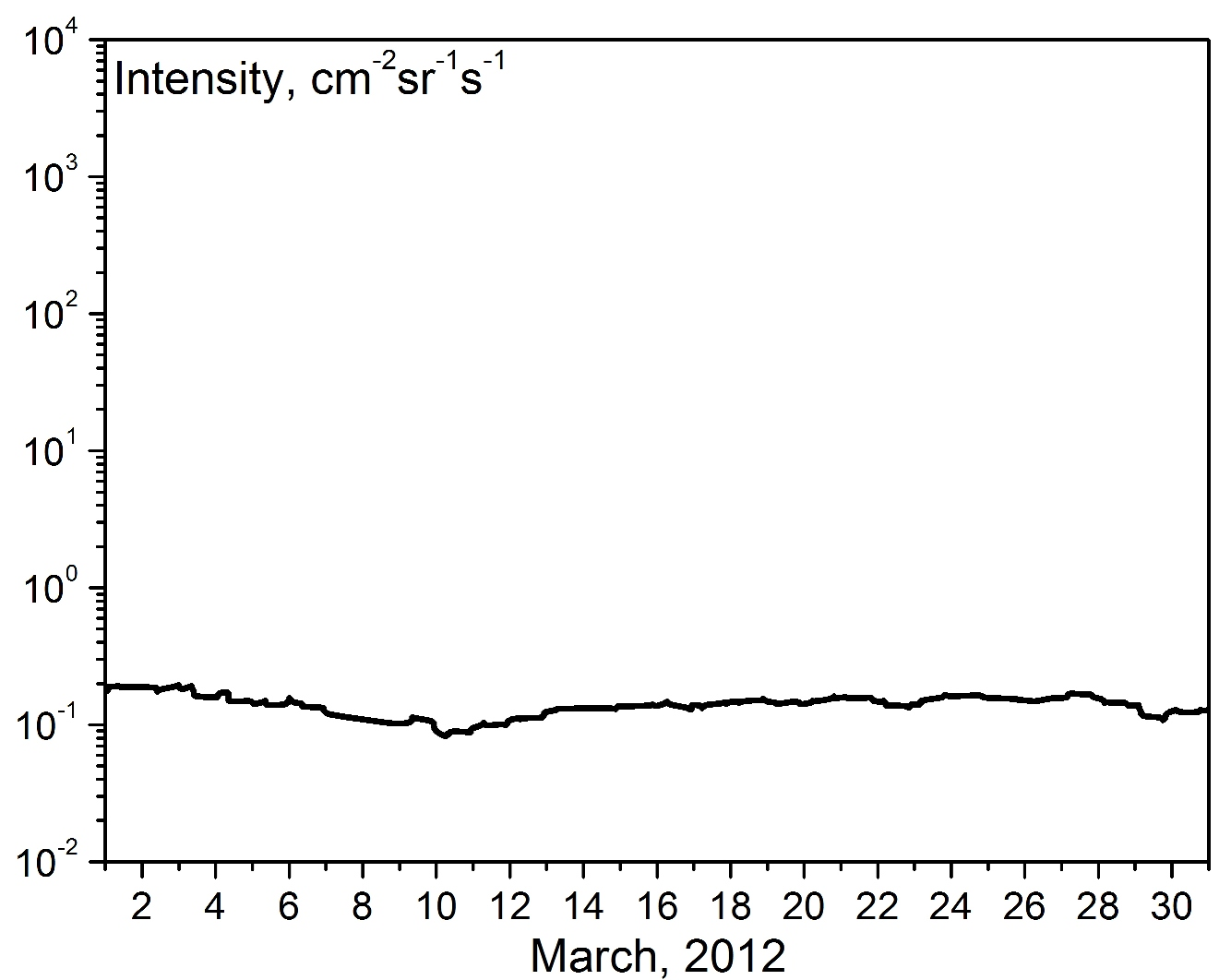}
\caption{4-6 MeV electron flux at L=2.5-2.7 (B$<$0.23) during March 2012.}
\label{fig2}
\end{figure}

\begin{figure} [h]
\includegraphics[width=85mm]{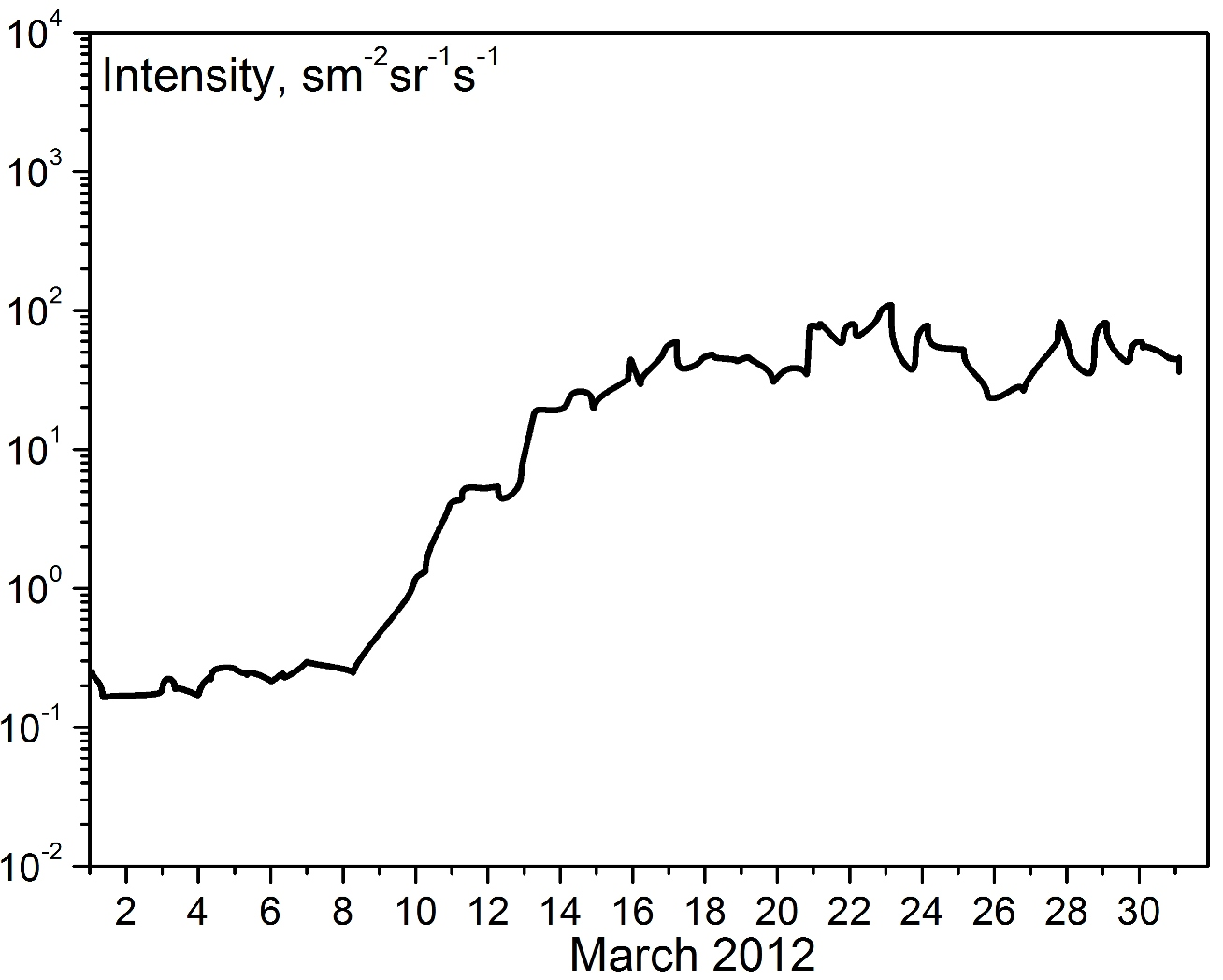}
\caption{4-6 MeV electron flux at L=2.9-3.1 (B$<$0.24) during March 2012.}
\label{fig3}
\end{figure}

\begin{figure} [h]
\includegraphics[width=73mm]{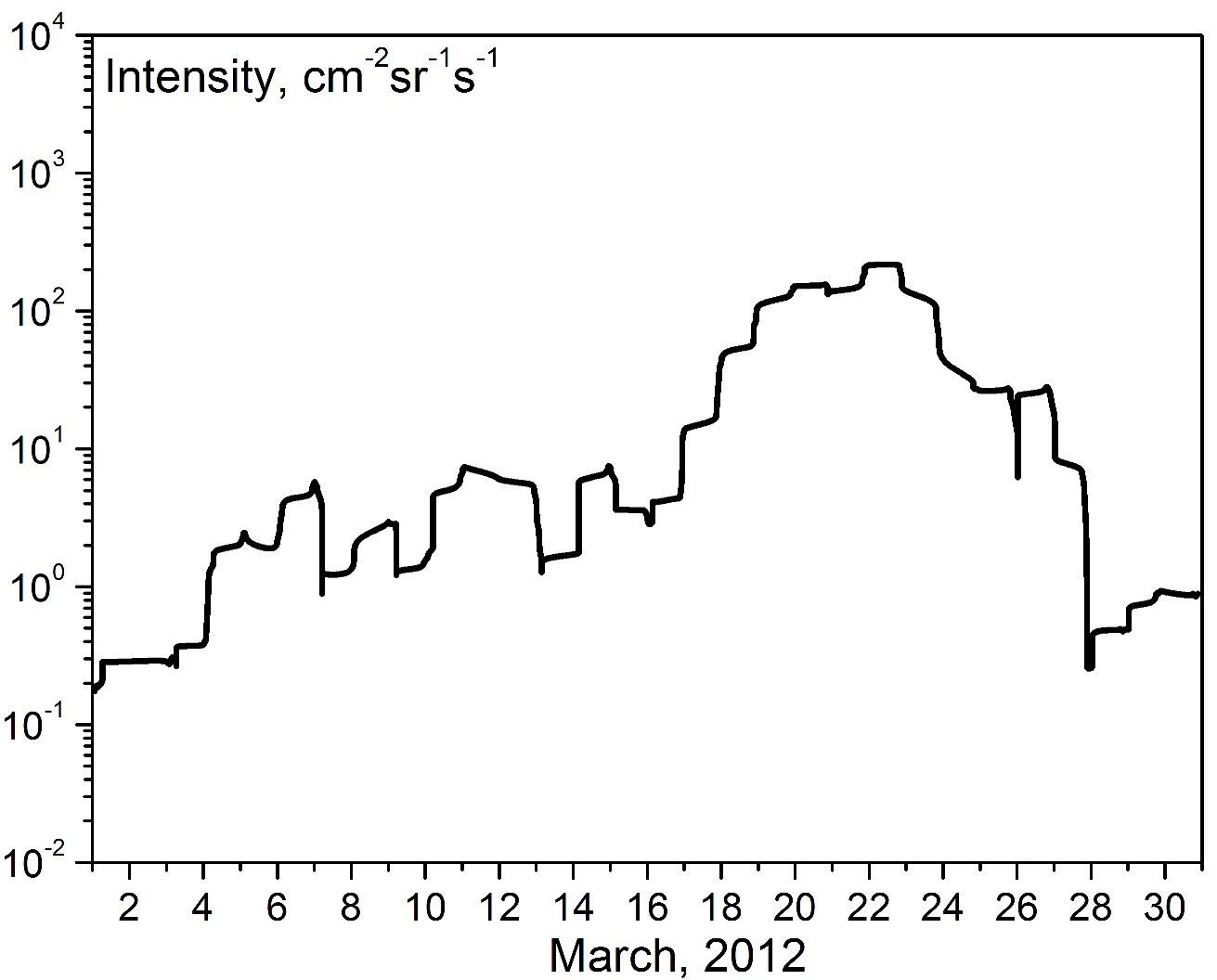}
\caption{4-6 MeV electron flux at L=4.5-5.0 (B$<$0.30) during March 2012.}
\label{fig4}
\end{figure}

\begin{figure}
\includegraphics[width=73mm]{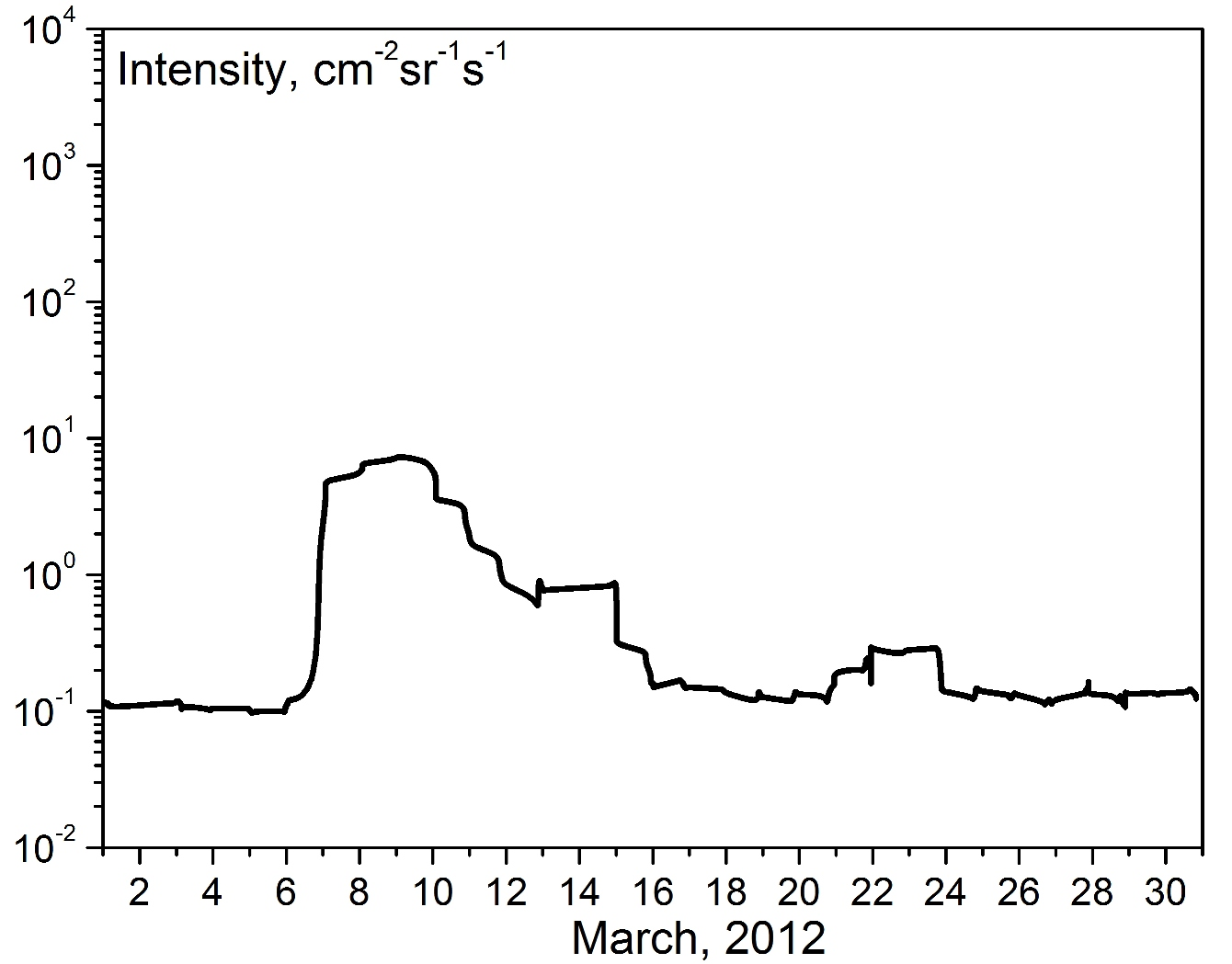}
\caption{4-6 MeV electron flux at L=8.0-9.0 (B$<$0.40) during March 2012.}
\label{fig5}
\end{figure}

\begin{figure} [h!]
\includegraphics[width=73mm]{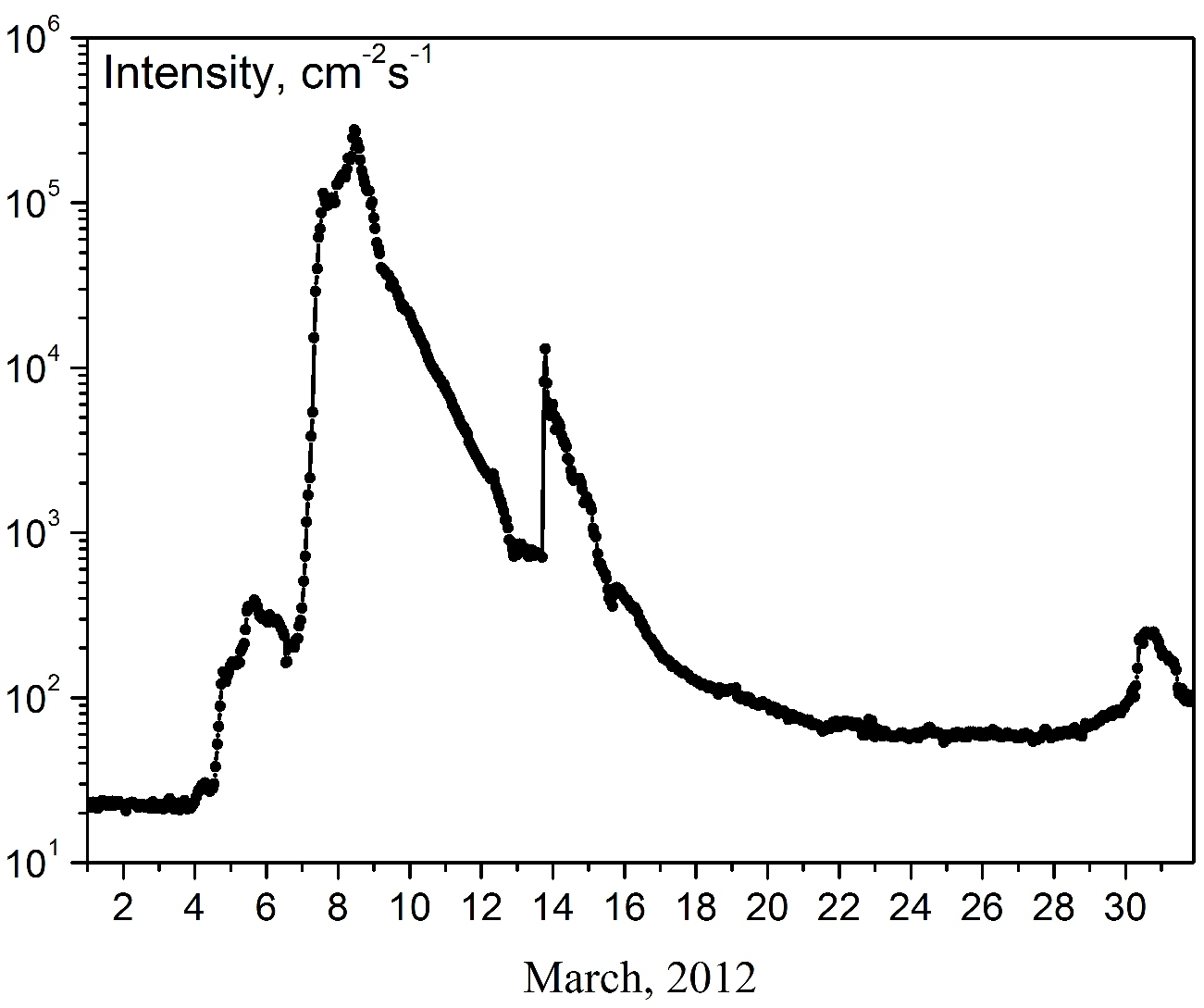}
\caption{ACE spacecraft data on electron flux (0.175-0.315 MeV) during March 2012.
https:$//$darts.isas.jaxa.jp/pub$/$sswdb$/$ace}
\label{fig6}
\end{figure}

Such behavior of high-energy electron flux was also observed in zone of outer radiation belt up to L=4.0.  In \cite{bib:bibl9} this high-energy component of electron belt has been analyzed during all 2012 year, and it was noted large and sharp variation of 4-6 MeV electron flux during strong geomagnetic storms (Dst $\sim$ -100 nT). In this work for slight and moderate disturbances (Dst$\sim$ -20 - -50 nT), as it was shown in Fig. 3, high-energy electron belt in the inner zone (L=2,9-4.0) is not sensitive and stays stable. 
In the outer zone (L$>$5) of the outer radiation belt high-energy component of electron flux becomes strongly variable and very sensitive to moderate level of Dst changes (Fig. 4) and possibly depends on the presence of MeV electrons on the outer boundary of the magnetosphere (Fig. 5 and Fig.6), because these electrons can penetrate into magnetosphere up to L=4-5 accelerating during the radial transfer.

Fig. 4 demonstrates complex profile of high-energy electron flux at L=4.5-5.0. Such shape of profile was also observed at L up to 7, that shows the influence of combination of two physics factors. First of them is the Dst variation (Fig. 1) and the second one is the presence of high flux of sub MeV electrons in interplanetary space (Fig. 6), generated in solar flare, and on the outer zone of the magnetosphere (Fig. 5).

% figures should be put into the text as floats.
% Use the graphics or graphicx packages (distributed with LaTeX2e)
% and the \includegraphics macro defined in those packages.
% See the LaTeX Graphics Companion by Michel Goosens, Sebastian Rahtz,
% entire page. There is no need to do explicit centering.

% \begin{figure}
% \end{figure}

% Surround figure environment with turnpage environment for landscape
% figure
% \begin{figure}
% \end{figure}

\section{Conclusion}
Analysis of ARINA experimental data on high-energy electrons, carried out in this work, showed the different behavior of electron fluxes in the outer radiation belt in dependence on L shells. In the inner zone (L=2.9-3.1) of the outer radiation belt the  high-energy electron flux variations are caused by large geomagnetic storms. In case of slight or moderate storms electron flux at this L is insensitive to geomagnetic disturbances. In the outer zone (L$>$4.5) of the outer radiation belt high-energy electron flux has the large variability defining by as moderate Dst changes as the presence of MeV electrons on the boundary of magnetosphere.

% If in two-column mode, this environment will change to single-column
% format so that long equations can be displayed. Use
%\begin{widetext}
%\end{widetext}

\bigskip % extra skip inserted
\begin{acknowledgments}

This work was supported by National Research Nuclear Universaty MEPhI in the framework of the Russian Academic Exellence Project (contract No.02.a03.21.0005. 27.08.2013).
\end{acknowledgments}

\bigskip % extra skip inserted

\end{document}